\documentclass[letterpaper,aps,prb,twocolumn,showpacs]{revtex4}
\usepackage{graphicx}
\usepackage{amsmath}
\usepackage{epstopdf}
\usepackage{amsfonts}
\usepackage{amssymb}

\begin{document}

\title{Current-induced magnetization dynamics in disordered itinerant
ferromagnets}
\author{Yaroslav Tserkovnyak}
\affiliation{Lyman Laboratory of Physics, Harvard University, Cambridge, Massachusetts
02138, USA}
\affiliation{Department of Physics and Astronomy, University of California, Los Angeles,
California 90095, USA}
\author{Hans Joakim Skadsem}
\affiliation{Department of Physics, Norwegian University of Science and Technology,
N-7491 Trondheim, Norway}
\author{Arne Brataas}
\affiliation{Department of Physics, Norwegian University of Science and Technology,
N-7491 Trondheim, Norway}
\affiliation{Centre for Advanced Study at the Norwegian Academy of Science and Letters,
Drammensveien 78, NO-0271 Oslo, Norway}
\author{Gerrit E. W. Bauer}
\affiliation{Kavli Institute of NanoScience, Delft University of Technology, 2628 CJ
Delft, The Netherlands}

\begin{abstract}
Current-driven magnetization dynamics in ferromagnetic metals are studied in
a self-consistent adiabatic local-density approximation in the presence of
spin-conserving and spin-dephasing impurity scattering. Based on a
quantum kinetic equation, we derive Gilbert damping and spin-transfer
torques entering the Landau-Lifshitz equation to linear
order in frequency and wave vector. Gilbert damping and a current-driven
dissipative torque scale identically and compete, with the result
that a steady current-driven domain-wall motion is insensitive to spin
dephasing in the limit of weak ferromagnetism. A uniform magnetization is
found to be much more stable against spin torques in the itinerant
than in the \textit{s}-\textit{d} model for ferromagnetism. A dynamic
spin-transfer torque reminiscent of the spin pumping in multilayers is identified and shown
to govern the current-induced domain-wall distortion.
\end{abstract}

\date{\today}
\maketitle


\section{Introduction}
\label{intro}

Metallic ferromagnets, notably the transition metals Fe, Co, and Ni, seem to
be well understood, at least at temperatures sufficiently below criticality.
Ground state properties such as cohesive energies, elastic constants,\cite
{Moruzzi:78} magnetic anisotropies in multilayers,\cite{Daalderop:prl92} but
also low-energy excitations that define Fermi surfaces,\cite{Choy:baps00}
spin-wave dispersions, and Curie temperatures\cite{Pajda:prb01} are computed
accurately and without adjustable parameters in the framework of local
spin-density--functional theory (SDFT).\cite{Kubler00} Transport properties
such as electric resistances due to random impurities are accessible to 
\textit{ab initio} band-structure calculations as well.\cite{Mertig:prb93}
However, important issues are still under discussion. Consensus has not been
reached, e.g., on the nature and modeling of the Gilbert damping of
the magnetization dynamics,\cite{Kunes:prb02,Koopmans:prl05} the anomalous
Hall effect,\cite{Yao:prl04} and the current-induced magnetization dynamics.
\cite
{Berger:prb96,Bazaliy:prb98,Fernandez:prb04,Tatara:prl04,Shibata:prl05,Li:prl04,Zhang:prl04,Li:jap05,Barnes:prl05}
The fundamental nature and technological importance of these effects make
them attractive research topics.

In this paper, we hope to contribute to a better understanding
of the interaction of an electric current with a magnetization order
parameter in dirty ferromagnets, motivated in part by the sophistication
with which the analogous systems of dirty superconductors has been mastered.
\cite{Belzig:sm99} To this end, we proceed from time-dependent SDFT in an
adiabatic local density approximation (ALDA) and the Keldysh Green's function
method in a quasiparticle approximation. We restrict ourselves to weak,
diffusive ferromagnets where spin dynamics take place near the Fermi
surface, an approximation that enables us to microscopically derive a simple
quantum kinetic equation for the electronic spin distribution. The kinetic
equation is used to derive a Landau-Lifshitz-Gilbert equation for the
spatiotemporal magnetization that significantly differs from earlier
phenomenological approaches based on the \textit{s}-\textit{d} model.
We apply the general theory to the current-driven spin-wave excitation and
domain-wall motion. After our paper was posted (cond-mat/0512715), Kohno~et~al.\cite{Kohno:condmat06} treated the
same problem by diagrammatic perturbation theory. For weak ferromagnets,
i.e., when the exchange potential is small compared to the Fermi energy, their
results agree with ours. For strong ferromagnets, they report small corrections.

The convincing evidence that transition-metal ground and weakly-excited
states are well described by the mean-field Stoner model provided by
local-SDFT can be rationalized by the strong hybridization between the
nearly free \textit{s}-\textit{p} bands and the localized \textit{d}
electrons.\cite{Kubler00} It implies that the orbital angular momentum is
completely quenched on time scales typical for the transport and
magnetization dynamics. Both electric current and magnetization are
therefore carried by the same itinerant Bloch states. The alternative 
\textit{s}-\textit{d} model, in which only the localized \textit{d}
electrons are intrinsically magnetic and affect the delocalized \textit{s}
electrons via a local spin-dependent exchange potential, is often used
because it is amenable to sophisticated many-body treatments. On a
mean-field level and with adjustable parameters, both models are completely
equivalent for static properties. We find that the magnetization dynamics
show drastic and experimentally testable differences that derive from the
necessity of a self-consistent treatment of the exchange potential in
itinerant ferromagnets that is not required in the \textit{s}-\textit{d} model.

The paper is organized as follows: In Sec.~\ref{model}, we discuss the model and the basic assumptions of the theory. In Sec.~\ref{qke}, the quantum kinetic equation is derived in the real-time Green's function formalism, which is then used to obtain the magnetic equation of motion in Sec.~\ref{eom}. The implications for the macroscopic dynamics are discussed in Sec.~\ref{cons}, before the paper is briefly summarized in Sec.~{\ref{sum}}.

\section{Model}
\label{model}

In time-dependent SDFT,\cite{Runge:prl84,Capelle:prl01,Qian:prl02} the
magnetic response is formally reduced to a one-body Hamiltonian in $2\times 2
$ Pauli spin space spanned by the unit matrix $\hat{1}$ and $\boldsymbol{
\hat{\sigma}}=(\hat{\sigma}_{x},\hat{\sigma}_{y},\hat{\sigma}_{z})$, the
vector of \textit{one-half} of the Pauli matrices:$\ $ 
\begin{align}
\mathcal{\hat{H}}& =\left[ \mathcal{H}_{0}+U(\mathbf{r},t)+V[\hat{\rho}](
\mathbf{r},t)\right] \hat{1}  \notag \\
& +\gamma \hbar \boldsymbol{\hat{\sigma}}\cdot \left( \mathbf{H}+\mathbf{H}_{
\mathrm{xc}}\left[ \hat{\rho}\right] \right) (\mathbf{r},t)+\mathcal{\hat{H}}
_{\sigma }\,,  \label{HKS}
\end{align}
where $\mathcal{H}_{0}$ is the crystal Hamiltonian, $U$ is the
scalar disorder potential including an external electric field, and $V$ the
spin-independent part of the exchange-correlation potential. We recognize on
the right-hand side the Zeeman energy due to the sum of externally-applied and
anisotropy magnetic fields $\mathbf{H}$ as well as an exchange-correlation
contribution $\mathbf{H}_{\mathrm{xc}}$, disregarding an
exchange-correlation magnetic field coupled to the orbital motion. Here, $
\gamma >0$ is (minus) the gyromagnetic ratio and, $\mathbf{H}_{\mathrm{xc}}$
and $V$ are functionals of the time-dependent spin-density matrix 
\begin{equation}
\rho _{\alpha \beta }(\mathbf{r},t)=\langle \Psi _{\beta }^{\dagger }(
\mathbf{r})\Psi _{\alpha }(\mathbf{r})\rangle _{t}
\end{equation}
that should be computed self-consistently from the Schr\"{o}dinger equation
corresponding to $\mathcal{\hat{H}}$. $\mathcal{\hat{H}}_{\sigma }$ is the
spin-nondiagonal Hamiltonian accompanying magnetic and spin-orbit
interaction potential disorder, thereby disregarding the \textquotedblleft
intrinsic\textquotedblright\ spin-orbit interaction in the bulk band
structure, apart from the crystal anisotropy contribution to $\mathbf{H}$.
Since we focus on low-energy magnetic fluctuations that are long range and
transverse, we may restrict our attention to a single band with effective
mass $m_{e}$. Systematic improvements for realistic band structures can be
made from this starting point. We furthermore adapt the ALDA form for the
exchange-correlation field: 
\begin{equation}
\gamma \hbar \mathbf{H}_{\text{xc}}[\hat{\rho}](\mathbf{r},t)\approx \Delta
_{\mathrm{xc}}\mathbf{m}(\mathbf{r},t)\,,
\end{equation}
where $\mathbf{m}$ is the local magnetization direction with $\left\vert 
\mathbf{m}\right\vert =1$ and $\Delta _{\mathrm{xc}}$ is the exchange
splitting averaged over the unit cell. In terms of the spin density 
\begin{equation}
\mathbf{s}(\mathbf{r})=\hbar \mbox{Tr}\left[ \boldsymbol{\hat{\sigma}}\hat{
\rho}(\mathbf{r})\right] \,,
\end{equation}
$\mathbf{m}=-\mathbf{s}/s_{0},$ where $s_{0}$ is the equilibrium value of $
\left\vert \mathbf{s}\right\vert $. For simplicity, the spin-independent
random component of the potential $U(\mathbf{r})$ is described as a
zero-average, Gaussian white noise correlator: 
\begin{equation}
\left\langle U(\mathbf{r})U(\mathbf{r^{\prime }})\right\rangle =\xi \delta (
\mathbf{r}-\mathbf{r^{\prime }})\,.  \label{U}
\end{equation}
A characteristic scattering time $\tau$ is defined by 
\begin{equation}
\xi=\frac{\hbar}{\pi(\nu _{\uparrow }+\nu _{\downarrow })\tau}\,,
\end{equation}
where $\nu _{s}$ is the spin-$s$ density of states at the Fermi level. We
consider two contributions to the spin-dephasing Hamiltonian $\mathcal{\hat{H
}}_{\sigma }$: spin-orbit scattering associated with the impurities and
scattering at magnetic disorder that is modeled as a static random exchange field $\mathbf{h}(\mathbf{r})$ with white-noise correlator
\begin{equation}
\left\langle h_\alpha(\mathbf{r})h_\beta(\mathbf{r^\prime})\right\rangle=\xi_\alpha\delta_{\alpha\beta}\delta(\mathbf{r}-\mathbf{r^\prime})\,.
\end{equation}
It turns out both can be captured in terms of a properly averaged, single
parameter $\tau _{\sigma}$ for the characteristic transverse spin-dephasing time in the equation of motion for the magnetization.
Derivation of the phenomenological $\tau _{\sigma }$ for concrete
microscopic models and dephasing mechanisms will be the topic of future
correspondence.

The ALDA is appropriate to describe corrections to the magnetization
dynamics linear in $\partial_{\mathbf{r}}$ that, although vanishing for
homogeneous systems,\cite{Capelle:prl01} are important in the presence of a
current bias. The second-order correction (in homogeneous isotropic systems)
is $H_{\text{ex}}^{\prime} \propto\boldsymbol{\hat{\sigma}}\cdot\partial_{
\mathbf{r}}^{2}\mathbf{m}$, which contributes to the spin-wave stiffness
[and can be taken into account via the effective field, see Eq.~(\ref{Heff})
below]. Not much is known about the importance of nonadiabatic many-body
corrections that in principle contribute to the magnetization damping.
However, for slowly-varying perturbations of a homogeneous ferromagnet in
time and space, the corrections to the ALDA are usually small.\cite
{Qian:prl02} Here we concentrate on dirty ferromagnets in which the impurity
(or phonon) scattering dominates quasiparticle scattering due to
electron-electron interactions.

In the next section, we derive the quantum kinetic equation for ferromagnetic dynamics by adiabatically turning
on a uniform electric field until a steady state is established for a given
current bias. The magnetization $\mathbf{m}$ is then perturbed with respect
to a uniform ground state configuration $\mathbf{m}_{0}=\mathbf{z}$. We then
compute small deviations of the spin density $\delta \mathbf{s}=\mathbf{s}
+s_{0}\mathbf{z}$, and replace $\mathbf{s}$ by $-s_{0}\mathbf{m}$ in the
resulting equations of motion, completing the self-consistency loop. A
natural approach to carry out these steps is the Keldysh Green's function
formalism, which we briefly outline in the following. If the reader is not interested in the technical details, we recommend jumping to Sec.~\ref{eom} for the discussion of the resulting equation of motion for the magnetization dynamics and Sec.~\ref{cons} for the physical consequences for macroscopic dynamics.

\section{Quantum kinetic equation}
\label{qke}

The Keldysh matrix Green's function can be represented by the retarded $\hat{G}
^{R}(x,x^{\prime })$, advanced $\hat{G}^{A}(x,x^{\prime })$, and Keldysh $
\hat{G}^{K}(x,x^{\prime })$ components,\cite{Rammer:rmp86} where $x$ denotes
position and time arguments. In the mixed (Wigner) representation $(\mathbf{r
},t;\mathbf{k},\varepsilon )$, in which $(\mathbf{r},t)$ are the center of
mass coordinates, and using the gradient approximation (valid when $\hbar
\partial _{t}\ll \Delta _{\mathrm{xc}}$ and $\partial _{r}\ll k_{F}$, a
characteristic Fermi wave number), the Keldysh component of the Dyson
equation reads in what is called the semiclassical approximation 
\begin{align}
\lbrack \hat{G}_{0}^{-1},\hat{G}^{K}]_{p}-& [\hat{G}^{K},\hat{G}
_{0}^{-1}]_{p}-2i[\hat{G}_{0}^{-1},\hat{G}^{K}]  \notag \\
& =\{\hat{\Sigma}^{K},\hat{A}\}-\{\hat{\Gamma},\hat{G}^{K}\}\,.  \label{Dp}
\end{align}
The left-hand side (l.h.s.) is the kinetic equation in the clean limit and the right-hand side (r.h.s.) is the
collision integral. In the derivation of this equation, self-energy
renormalization effects on the l.h.s and gradient corrections to the
collision integral have been disregarded. This requires that $\Delta_{\mathrm{xc}}/\mu \ll 1$, where $\mu$ is the Fermi energy, although the corrections for large $\Delta_{\rm xc}$ appear to be very small, see below. $\hat{\Sigma}$ is the self-energy due
to disorder, which has three nontrivial components ($R$, $A$, and $K$) along
the Keldysh contour. Here, 
\begin{equation}
\lbrack \hat{B},\hat{C}]_{p}=\partial _{x}\hat{B}\cdot \partial _{p}\hat{C}
-\partial _{p}\hat{B}\cdot \partial _{x}\hat{C}
\end{equation}
is the generalized Poisson bracket (where $\partial _{x}\cdot \partial
_{p}=\partial _{\mathbf{r}}\cdot \partial _{\mathbf{k}}-\hbar \partial
_{t}\partial _{\varepsilon }$), $[,]$ and $\{,\}$ are matrix commutators and
anticommutators,
\begin{equation}
\hat{A}=i(\hat{G}^{R}-\hat{G}^{A})
\end{equation}
and
\begin{equation}
\hat{\Gamma}=i(\hat{\Sigma}^{R}-\hat{\Sigma}^{A})\,.
\end{equation}
$\hat{G}_{0}^{-1}$ is the inverse of
the (retarded or advanced) Green's function in the absence of disorder: 
\begin{equation}
\hat{G}_{0}^{-1}(\mathbf{r},t;\mathbf{k},\varepsilon )=[\varepsilon
-\varepsilon _{{k}}+e\varphi (\mathbf{r},t)]{\hat{1}}-\Delta _{\mathrm{xc}}
\boldsymbol{\hat{\sigma}}\cdot \mathbf{m}(\mathbf{r},t)\,,  \label{G0}
\end{equation}
where $\varphi $ is the potential due to an applied electric field, and
\begin{equation}
\varepsilon_{{k}}=\frac{(\hbar k)^{2}}{2m_{e}}-\mu
\end{equation}
are the eigenvalues of $
\mathcal{H}_{0}$. We have disregarded the magnetic field for the moment. In
the self-consistent Born approximation for scalar disorder scattering, the
self-energy becomes 
\begin{equation}
\hat{\Sigma}(\mathbf{r},t;\mathbf{k},\varepsilon )=\xi \int dk^{\prime }
\hat{G}(\mathbf{r},t;\mathbf{k^{\prime }},\varepsilon )
\end{equation}
for each of the three components, where $dk^{\prime }=d^{3}\mathbf{k}
^{\prime }/(2\pi )^{3}$. Self-energies for spin-dependent scattering
channels can be calculated analogously. For $\Delta _{\mathrm{xc}}/\mu \ll 1$
, we approximate the spectral function by Dirac delta functions at the two
spin bands. Note that even though we are considering weak ferromagnets, the
impurity concentration is still considered dilute, so that $\hbar /\tau
,\hbar /\tau _{\sigma }\ll \Delta _{\mathrm{xc}},\mu $. By
disregarding gradient terms of self-energies and the spectral function in
the derivation of Eq.~(\ref{Dp}), the Wigner representation
transformed the collision integral into a local form. Gradient corrections
disappear when the system is spatiotemporally homogeneous and/or we restrict
our attention to weak ferromagnets, thereby discarding corrections of order $
\mathcal{O}\left( (\hbar /\tau ,\Delta _{\mathrm{xc}})/\mu \right) $. In
spite of this restriction, we believe that our formalism still captures the
essential physics of the model (and therefore transition-metal
ferromagnets) in a clear and coherent fashion. Assessing the
leading corrections to our treatment would require one to reconsider as well the
simple ALDA mean-field treatment we are relying on.

We concentrate now on the spin dynamics for small deviations of the
magnetization direction $\mathbf{m}=\mathbf{z}+\mathbf{u}$ from the $z$ axis
($\mathbf{u}\perp \mathbf{z}$) in the presence of a weak uniform electric
field $\mathbf{E}=-\partial _{\mathbf{r}}\varphi $ in the quasiparticle
approximation for the Keldysh Green's function, 
\begin{equation}
\hat{G}^{K}(\mathbf{r},t;\mathbf{k},\varepsilon )=-2\pi i\sum_{s}\delta
(\varepsilon -\varepsilon _{ks})\hat{g}_{\mathbf{k}s}(\mathbf{r},t)\,,
\label{qp}
\end{equation}
where
\begin{equation}
\varepsilon _{ks}=\varepsilon _{k}+\frac{s}{2}\Delta _{\mathrm{xc}}\,.
\end{equation}
Two
spin bands labeled by $s=\uparrow ,\downarrow =\pm $ become separated when
the disorder is weak. Note that in equilibrium,
\begin{equation}
\hat{g}_{\mathbf{k}s}=\left(\frac{1}{2}+s\hat{\sigma}_{z}\right)\tanh\left(\frac{\varepsilon _{ks}}{2k_{B}T}\right)\,,
\end{equation}
where $T$ is the temperature and $k_{B}$ the
Boltzmann constant. The electric field applied to a rigidly-uniform
ferromagnet, $\mathbf{u}=0,$ excites a nonequilibrium distribution $\hat{g}_{
\mathbf{k}s}$ that is also diagonal in the spin indices. Interband spin-flip
scattering vanishes upon momentum integration, since a weak uniform electric
field induces only a $p$-wave distribution. The transport in each spin band
(obtained by integrating Eq.~(\ref{Dp}) over energy $\varepsilon $ at fixed $
\mathbf{k}$ near $\varepsilon _{ks}$) is thus described by the conventional
Boltzmann equation,\cite{Valet:prb93} at $T\rightarrow 0$ solved by the
\textquotedblleft drift\textquotedblright\ distribution 
\begin{equation}
\delta \hat{g}_{\mathbf{k}s}=\frac{\hbar e}{\pi \xi \nu _{s}}\left(\frac{1}{2}+s\hat{\sigma}_{z}\right)\mathbf{E}\cdot\mathbf{v}_{\mathbf{k}}\delta (\varepsilon _{ks})\,.
\end{equation}

The distribution functions $\hat{g}_{\mathbf{k}s}$ acquire off-diagonal
components (describing transverse spins) in the presence of a finite $
\mathbf{u}$ (so that out of equilibrium the spin subscript should not be
taken literally). Equation (\ref{Dp}) leads to the linearized kinetic equation
for the transverse component $\hat{g}_{\mathbf{k}s}^{T}=\mathbf{g} _{\mathbf{
k}s}\cdot\boldsymbol{\hat{\sigma}}$ ($\mathbf{g}_{\mathbf{k}s} \perp\mathbf{z
}$): 
\begin{align}
& \hbar\partial_{t} \mathbf{g}_{\mathbf{k}s} + \hbar(\mathbf{v}_{\mathbf{k}}
\cdot\partial_{\mathbf{r}})\left[ \mathbf{g}_{\mathbf{k}s}-\Delta_{\mathrm{xc
} }\mathbf{u}\delta(\varepsilon_{ks})\right] - \Delta_{\mathrm{xc}}\mathbf{z}
\times\mathbf{g}_{\mathbf{k}s}  \notag \\
&+s\Delta_{\mathrm{xc}}\mathbf{z} \times \mathbf{u}\,\mathrm{sign}
(\varepsilon_{ks}) + \frac{s\hbar e}{\pi\xi\nu_{s}}\left(\mathbf{E}\cdot
\mathbf{v}_{\mathbf{k}}\right) \Delta_{\mathrm{xc}}\mathbf{z}\times\mathbf{u}
\,\delta(\varepsilon_{ks})  \notag \\
&-e\left(\mathbf{E}\cdot\partial_{\mathbf{k}}\right) \mathbf{g}_{\mathbf{k}
s}=\pi\xi\sum_{s^{\prime}}\int dk^{\prime}\:\delta(\varepsilon_{k^\prime
s^\prime} - \varepsilon_{ks})\left[ \mathbf{g}_{\mathbf{k}^\prime s^\prime}
- \mathbf{g}_{\mathbf{k}s} \right.  \notag \\
&\left. + (s-s^\prime)\mathbf{u}\,\mathrm{sign}(\varepsilon_{ks})\right] +
(\nu_{-s}/\nu_{s}-1) \hbar e\left(\mathbf{E}\cdot\mathbf{v}_{\mathbf{k}
}\right) \mathbf{u}\delta(\varepsilon_{ks})  \notag \\
&-\frac{\hbar}{\tau_{\sigma}}\left(\mathbf{g}_{\mathbf{k}s} - s\mathbf{u}
\left[ \mathrm{sign}(\varepsilon_{ks})+\frac{\hbar e}{\pi\xi\nu_{s}} \mathbf{
E} \cdot \mathbf{v}_{\mathbf{k}} \delta(\varepsilon_{ks}) \right]\right)
\hspace{-0.01cm}.  \label{kH}
\end{align}
Quasiparticles propagate with group velocity $\mathbf{v}_{\mathbf{k}
}=\partial_{\mathbf{k}}\varepsilon_{k}/\hbar$. On the l.h.s., an
inhomogeneous exchange field is seen to cause electron acceleration and spin
precession. The second term on the second line describes spin precession of
electrons accelerated by the electric field and the following term
acceleration of the precessed electrons. On the r.h.s. we recognize elastic
disorder scattering and transverse spin relaxation, the latter in terms of
the spin-dephasing time $\tau_{\sigma}$. Energy-conserving mixing between
the spin bands is allowed by disorder (in the presence of transverse
fields), as reflected in the $s^{\prime}=-s$ part of the collision integral.
We also took into account the contribution to the r.h.s.~of Eq.~(\ref{Dp})
from anticommuting the current-induced drift Keldysh component with the
spectral-function correction due to the magnetization deviation $\mathbf{u}$:
\begin{equation}
\delta\hat{A} = 2\pi\boldsymbol{\hat{\sigma}} \cdot \mathbf{u} \sum_{s} s
\delta(\varepsilon-\varepsilon_{ks})\,.
\end{equation}

\section{Magnetic equation of motion}
\label{eom}

Integrating the kinetic equation (\ref{kH}) over momentum yields the
equation of motion for the nonequilibrium spin density $\delta \mathbf{s}
=-(\hbar /4)\sum_{s}\int dk\mathbf{g}_{\mathbf{k}s}$: 
\begin{align}
\partial _{t}\delta \mathbf{s}& -\frac{\Delta_{\mathrm{xc}}}{\hbar}\mathbf{z}
\times \delta \mathbf{s}-\frac{\Delta_{\mathrm{xc}}}{\hbar}\mathbf{z}\times 
\mathbf{u}s_{0}  \notag \\
& =\frac{\hbar}{4}\sum_{s}\int dk(\mathbf{v}_{\mathbf{k}}\cdot \partial _{\mathbf{
r}})\mathbf{g}_{\mathbf{k}s}-\frac{\delta \mathbf{s}+\mathbf{u}s_{0}}{\tau_{\sigma}}\,.  \label{ds}
\end{align}
The integral on the r.h.s. is the divergence of the spin-current density,
determined by the $p$-wave component of $\mathbf{g}_{\mathbf{k}s}$, which
can be found by a tedious (but straightforward) manipulation of the kinetic
equation. Confining our interest to spatially slowly varying phenomena
results in a major simplification: since $\partial _{\mathbf{r}}$ already
appears in Eq.~(\ref{ds}), we can disregard spatial derivatives in the $p$-wave component of $\mathbf{g}_{\mathbf{k}s}$. We can now also include a
static field $H\ll \Delta _{\mathrm{xc}}$ along the $z$-axis by substituting
primed quantities $\Delta _{\mathrm{xc}}^{\prime }=\Delta _{\mathrm{xc}
}+\gamma H$ and $\mathbf{u}^{\prime }=-(1-\gamma H/\Delta _{\mathrm{xc}
}^{\prime })\delta \mathbf{s}/s_{0}$ for the corresponding unprimed ones in
the above expressions. The final result for the small-angle transverse spin
dynamics is 
\begin{equation}
\partial _{t}\mathbf{u}=\omega _{0}\mathbf{z}\times \mathbf{u}-\beta \omega
_{0}\mathbf{u}+\mathcal{P}\left[ 1-\mathbf{z}\times \frac{\hbar \partial _{t}
}{\Delta _{\mathrm{xc}}}\right] \left( \mathbf{j}\cdot \partial _{\mathbf{r}
}\right) \mathbf{u}\,,  \label{eou}
\end{equation}
disregarding the $\mathcal{O}(1/\Delta _{\mathrm{xc}}^{2})$ terms inside the
square brackets. Here $\mathbf{j}$ is the applied current density bias, $
\gamma \mathbf{H}=\omega _{0}\mathbf{z}$,
\begin{equation}
\beta =\frac{\hbar}{\tau _{\sigma}\Delta _{\mathrm{xc}}}\,,
\end{equation}
and $\mathcal{P}=(\hbar /2e)P/s_{0}$, where $
P=(\sigma _{\uparrow }-\sigma _{\downarrow })/(\sigma _{\uparrow }+\sigma
_{\downarrow })$ is the conductivity spin polarization, $\sigma _{s}$ being
the conductivity for spin $s$ along $-\mathbf{m}$. For a Drude conductivity
of parabolic bands, $P=\Delta _{\mathrm{xc}}/(\varepsilon _{F\uparrow
}+\varepsilon _{F\downarrow })$. We can transform the Bloch-like damping
term in Eq.~(\ref{eou}) to the Gilbert form by multiplying the equation by $
1-\beta \mathbf{z}\times$ from the left, which brings us to our
central result: 
\begin{equation}
\partial _{t}\mathbf{m}=\left. \partial _{t}\mathbf{m}\right\vert _{\text{
\textrm{LLG}}}+\left. \partial _{t}\mathbf{m}\right\vert _{j}\,,
\label{dtm}
\end{equation}
where 
\begin{equation}
\left. \partial _{t}\mathbf{m}\right\vert _{\text{\textrm{LLG}}}=-\gamma 
\mathbf{m}\times \mathbf{H}+\beta \mathbf{m}\times \partial _{t}\mathbf{m}
\label{llg}
\end{equation}
is the usual Landau-Lifshitz-Gilbert (LLG) with Gilbert damping
\begin{equation}
\alpha_{\mathrm{LDA}}=\beta\,.
\end{equation} 
\begin{equation}
\left. \partial _{t}\mathbf{m}\right\vert _{j}=\mathcal{P}\left[ 1-\mathbf{m}
\times \left( \beta +\frac{\hbar \partial _{t}}{\Delta _{\mathrm{xc}}}
\right) \right] \left( \mathbf{j}\cdot \partial _{\mathbf{r}}\right) \mathbf{
m}\,,  \label{mjc}
\end{equation}
where, as before, we neglect the $\mathcal{O}(1/\Delta _{\mathrm{xc}}^{2})$
terms. $\alpha_{\mathrm{LDA}}=\hbar/\tau _{\sigma }\Delta _{\mathrm{
xc}}$ relates the collective magnetization damping to the
single-electron spin relaxation that can be measured independently\cite
{Koopmans:prl05} and is consistent with experiments in permalloy films.\cite
{Ingvarsson:prb02} 
 Equations (\ref{llg}) and (\ref{mjc}) hold for small deviations from a
 homogeneous equilibrium state, but have the correct
 spin-rotationally invariant form valid also for long-wavelength
 large-angle dynamics when the magnetic state is locally close to the
 equilibrium configuration (which requires a large exchange splitting
 in comparison with other relevant energy scales). In particular,
 Eq.~(\ref{mjc}) should correctly describe domain walls wider and
 spin-wave lengths longer than the magnetic coherence length $\hbar v_F/\Delta_{\rm xc}$. For the same reason, the field $\mathbf{H}$ does not have to be nearly collinear with $\mathbf{m}$.

We can apply our method also to the mean-field \textit{s}-\textit{d} model
\cite{Tserkovnyak:apl04} which leads to interesting differences. We
reproduced the phenomenologically derived Eq.~(11) of Ref.~\onlinecite{Zhang:prl04} (plus the dynamic term linear in $\partial_{t}$). The Gilbert damping becomes reduced by the fraction $\eta$ of the total
spin angular momentum carried by the \textit{s} electrons, while $\beta$ is unmodified:
\begin{equation}
\alpha_{s-d}=\eta\beta\,,
\end{equation}
assuming $\eta\ll1$ (Ref.~\onlinecite{Tserkovnyak:apl04}). We
will see in the following that the ratio $\beta/\alpha$ determines several
interesting physical quantities with $\beta/\alpha_{\mathrm{LDA}}=1$ being a
very special point. A sizable \textit{s}-\textit{d} character of the
ferromagnetism alters this ratio, which could also be affected by a possible 
\textit{d}-magnetization damping in addition to the \textit{s}-electron
dephasing treated here.

Soon after our paper was posted (cond-mat/0512715), a diagrammatic treatment of spin torques in static weakly disordered localized and
itinerant ferromagnets has been reported by Kohno~\textit{et~al.}\cite
{Kohno:condmat06} Their calculation is not restricted to weak ferromagnets [although it misses dynamic current-driven torques such as the last term in Eq.~(\ref{mjc})],
and they find that in contrast to our result $\beta$ is not universally
identical to $\alpha$ in the LDA approximation. However, the ratio $\beta/\alpha_{\rm LDA}$, in Ref.~\onlinecite{Kohno:condmat06} expressed by the ratio
between the density of states averaged over the two Fermi surfaces and the
energy range spanned by $\Delta_{\mathrm{xc}}$, is close to unity for
almost all systems of interest. In particular, at low temperatures
and in three dimensions, Kohno~\textit{et~al.}'s expressions can be
evaluated to be $\beta/\alpha_{\rm LDA}\approx 1+(1/48)(\Delta_{\mathrm{xc}}/\mu)^{2}$ (with the same correction for the $\beta/\alpha_{s-d}$ ratio). This quadratic deviation from unity is very small; even for
$\Delta _{\mathrm{xc}}/\mu \sim 1/2$ it only amounts to about half a
percent. The present quasiparticle treatment is not well suited to study
ferromagnets exhibiting an arbitrarily strong exchange splitting (due to the
increasing importance of gradient corrections to the semiclassical
approximation with stronger exchange splittings). We are therefore hesitant
to make predictions for half-metallic ferromagnets. We are, however, confident that we
capture the important physics of most experimental systems to date. For this
reason, the present framework can also be used in studies of, e.g., relevant
spin-dephasing mechanisms and microscopically derived scattering rates. The
influence of realistic band structure effects, intrinsic spin-orbit, and
Coulomb interaction, as well as corrections beyond the mean-field
description might be more important than the gradient corrections to the
ratio $\beta/\alpha$. Furthermore, it is in general possible that other than impurity-related dephasing processes may contribute differently to $\alpha$ and $\beta$, especially in the presence of strong anisotropies.

\section{Current-driven domain-wall motion and bulk instabilities}
\label{cons}

Let us proceed by discussing the influence of $\beta /\alpha $ on the
magnetization dynamics, and in particular the limiting case in which this
ratio is unity. The dominant term $\boldsymbol{\tau }=\mathcal{P}\left( 
\mathbf{j}\cdot \partial _{\mathbf{r}}\right) \mathbf{m}$ in Eq.~(\ref{mjc})
is the conventional spin-transfer torque that, as far as the equation of
motion is concerned, can be absorbed into the magnetic free energy.\cite
{Bazaliy:prb98,Fernandez:prb04,Tatara:prl04} The (dissipative) term
proportional to $\beta $ acts like a magnetic field parallel to the
direction of the magnetization gradient in the current direction. This term
appears in our treatment by transforming Eq.~(\ref{eou}) into the LLG form (\ref{llg}). Zhang and Li\cite{Zhang:prl04} noted that although
this \textquotedblleft effective field\textquotedblright\ is much smaller
than $\boldsymbol{\tau}$ when $\beta \ll 1$, it has a qualitative effect on
the domain-wall motion. For example, in the absence of an external
magnetic field, a finite terminal velocity of a current-driven N\'{e}el wall
is found for all currents only when the effective field does not vanish.
Judging from the importance of dynamic corrections to the spin torques in
multilayer structures,\cite{Tserkovnyak:rmp05} the dynamic contribution in
Eq.~(\ref{mjc}) could be as significant since the typical frequencies of
ferromagnetic dynamics are $\omega \sim \tau _{\sigma }^{-1}$.

In this section, we discuss several experimental consequences
for $\mathbf{j}=j\mathbf{z}$, and a net effective field 
\begin{equation}
\mathbf{H}=\left( Km_{z}+H\right) \mathbf{z}-K_{\perp }m_{x}\mathbf{x}
+A\nabla ^{2}\mathbf{m}\,.  \label{Heff}
\end{equation}
Here, $K$ is an easy axis and $K_{\perp }$ an easy-plane anisotropy
constant, $A$ is the exchange-stiffness, and $H$ is the applied magnetic
field. $K,~K_{\perp },~A,~H\geq 0$.

Let us first consider current-driven domain-wall motion in the absence of applied field,  $H=0$. At the onset of the applied current density, a N\'{e}el wall
along the $z$ direction of width $W=\sqrt{A/K}$ with magnetization in the $
yz $ plane (pointing along $z$ at $z\rightarrow -\infty $ and in the
opposite direction at $z\rightarrow \infty $) starts to move\cite
{Zhang:prl04} with velocity (for not too large currents)
\begin{equation}
v_{i}=-\mathcal{P}j\,,
\label{vi}
\end{equation}
acquiring a terminal steady velocity for a constant current density given by
\begin{equation}
v_{f}=-\frac{\beta}{\alpha}\mathcal{P}j\,.
\label{vf}
\end{equation}
We find that the terminal velocity (\ref{vf}) is not influenced by the dynamic term on the r.h.s. of Eq.~(\ref{mjc}), and we get $v_{f}/v_{i}=\beta /\alpha _{\mathrm{LDA}}=1$ for the self-consistent LDA model of itinerant ferromagnetism. The initial velocity (\ref{vi}) agrees with expectations based on
angular-momentum conservation, and, curiously, for our model, the terminal
velocity is the same. According to Ref.~\onlinecite{Kohno:condmat06}, in
three dimensions, the correction to $\beta/\alpha_{\rm LDA}$ of order $(\Delta _{\mathrm{xc}}/\mu )^{2}$
is positive, which means that $v_{f}\gtrsim v_{i}$. Yamaguchi~\textit{et~al.}
\cite{Yamaguchi:prl04} expressed the current-induced domain wall velocity
\begin{equation}
v_{f}=-\zeta \mathcal{P}j
\end{equation}
in terms of an \textquotedblleft
efficiency\textquotedblright\ $\zeta $ of spin-current conversion into magnetization
dynamics. Their experimental value $\zeta \sim 0.1$ is much smaller than our
result of $\zeta =1$ in the absence of bulk or interface pinning (which, if
smooth enough, could in principle be added to the effective field $\mathbf{H}
$). For currents in excess of a threshold imposed by extrinsic pinning
defects, Barnes and Maekawa\cite{Barnes:prl05} predicted $\zeta =1$ for an 
\textit{s}-\textit{d} model, in contrast to a nonuniversal mean-field result 
$\zeta =\beta /\alpha _{s-d}=1/\eta$ of Ref.~\onlinecite{Zhang:prl04} which we confirm here.

Under the action of the current-induced spin torque, the shape of the moving domain wall distorts somewhat with respect to the equilibrium configuration. The corresponding domain-width change from the equilibrium value $W$ to the steady-state value $W_f$ was calculated in Ref.~\onlinecite{Li:prl04} using the Walker's ansatz. After generalizing their method to include the effects of $\beta$ as well as the dynamic term in the magnetic equation of motion (\ref{mjc}), we find
\begin{equation}
1-\frac{W_{f}}{W}\approx \frac{(\mathcal{P}j)^{2}}{2\gamma A}\left[ \frac{1}{
\gamma K_{\perp }}\left( 1-\frac{\beta }{\alpha }\right) ^{2}-\frac{\hbar }{
\Delta _{\mathrm{xc}}}\frac{\beta }{\alpha }\right] \,,  \label{W}
\end{equation}
where the first (second) term on the r.h.s. describes the wall deformation
due to the static (dynamic) part of Eq.~(\ref{mjc}). Now, considering $
\alpha _{\mathrm{LDA}}=\beta $, the first term vanishes and the wall
slightly broadens, unlike the wall compression predicted for the \textit{s}-\textit{d} model with a finite damping $\alpha$ but setting $\beta=0$ (Ref.~\onlinecite{Li:prl04}).

Finally, we discuss small-amplitude spin-wave solutions of Eqs.~(\ref{dtm}), (\ref{llg}), and (\ref{mjc}) of the form 
\begin{equation}
\mathbf{m}(\mathbf{r},t)=\mathbf{z}+\mathbf{u}_{0}\exp [i(\mathbf{q}\cdot\mathbf{r}-\omega t)]\,.
\end{equation}
We are especially interested in solutions with $\mbox{Im}\omega>0$, which describe exponentially growing spin-wave amplitude, signaling the onset of current-driven instabilities. We find that the critical current corresponding to $\mbox{Im}\omega=0$ is
determined from 
\begin{equation}
b^{\prime 2}\left(1-\frac{\beta}{\alpha}\right) ^{2}=\left( H^{\prime }+\frac{\beta}{\alpha}\frac{b^{\prime 2}\hbar}{\Delta_{\mathrm{xc}}}\right)\left( K^{\prime }+H^{\prime }+\frac{\beta}{\alpha}\frac{b^{\prime 2}\hbar}{\Delta_{\mathrm{xc}}}\right) \,,  \label{bc}
\end{equation}
where $b^{\prime }=\mathcal{P}(\mathbf{q}\cdot \mathbf{j})$, $H^{\prime
}=\gamma (H+K+Aq^{2})$, and $K^{\prime }=\gamma K_{\perp }$. For $\beta\to0$,
this reduces to
\begin{equation}
|b^{\prime }|\to\sqrt{H^{\prime }(K^{\prime }+H^{\prime })}
\label{doppler}
\end{equation}
which can be thought of as the Doppler shift due to drifting spins necessary
to overcome the natural spin-wave frequency.\cite
{Bazaliy:prb98,Fernandez:prb04,Li:prl04,Shibata:prl05} Our result that $
\alpha _{\mathrm{LDA}}=\beta $ for weak ferromagnets, however, implies that
a uniform magnetic state is stable against current-driven torques. In
general, the critical current density $j_{c}$ determined from Eq.~(\ref{bc})
can be significantly enhanced (depending on how close $\alpha$ and $\beta$ are) with respect to the \textquotedblleft Doppler-shift value" $j_{c0}$ calculated from Eq.~(\ref{doppler}):
\begin{equation}
j_{c}=\frac{j_{c0}}{|1-\beta /\alpha |}\,,
\end{equation}
where small corrections proportional to $\beta$ on the r.h.s. of
Eq.~(\ref{bc}) have been disregarded.

\section{Summary}
\label{sum}

In conclusion, we have used a quasiparticle approximation, valid for weak
ferromagnets, to derive an equation of motion for the magnetization dynamics
of disordered ferromagnets similar to the conventional LLG equation (\ref{llg}) with Gilbert damping $\alpha$ and a current-induced contribution
(\ref{mjc}) that is parametrized by a normalized single-electron spin-dephasing rate $
\beta =\hbar /\tau _{\sigma }\Delta _{\mathrm{xc}}$. By virtue of the
quasiparticle approximation, we obtain intuitively appealing kinetic
equations that clearly reflect the physical processes involved.

Within a self-consistent picture based on the local density approximation, we related the macroscopic damping in weak itinerant ferromagnets to the microscopic spin dephasing: $\alpha_{\mathrm{LDA}}=\beta$, and pointed out striking implications for current-driven macroscopic dynamics when the ratio $\beta/\alpha$ is close to unity (which can also be expected for strong ferromagnets in the ALDA approximation). We furthermore noted remarkable differences in the dynamics of itinerant ferromagnets, supposedly well-described by the local-density approximation, and those with localized \textit{d} or \textit{f} electron magnetic moments.

\acknowledgments

This work was supported in part by the Harvard Society of Fellows, the
Research Council of Norway through Grant Nos. 158518/143, 158547/431, and
167498/V30, and the National Science Foundation Grant No. PHY~99-07949.


\begin{thebibliography}
\expandafter\ifx\csname
natexlab\endcsname\relax

\fi
\expandafter\ifx\csname bibnamefont\endcsname\relax

\fi \expandafter\ifx\csname
bibfnamefont\endcsname\relax 

\fi
\expandafter\ifx\csname citenamefont\endcsname\relax

\fi \expandafter\ifx\csname
url\endcsname\relax 

\fi
\expandafter\ifx\csname urlprefix\endcsname\relax

\fi \providecommand{\bibinfo}[2]{#2} \providecommand{\eprint}[2][]{\url{#2}}

\bibitem[Moruzzi et~al.(1978)Moruzzi, Janak, and Williams]{Moruzzi:78} 
\bibinfo{author}{\bibfnamefont{V.~L.} \bibnamefont{Moruzzi}}, 
\bibinfo{author}{\bibfnamefont{J.~F.} \bibnamefont{Janak}}, and 
\bibinfo{author}{\bibfnamefont{A.~R.}
    \bibnamefont{Williams}}, \emph{
\bibinfo{title}{Calculated
      electronic properties of metals}}
(\bibinfo{publisher}{Pergamon
    Press}, \bibinfo{year}{1978}), \bibinfo{edition}{8th} ed.

\bibitem[Daalderop et~al.(1992)Daalderop, Kelly, and Broeder]
{Daalderop:prl92} 
\bibinfo{author}{\bibfnamefont{G.~H.~O.}
\bibnamefont{Daalderop}}, 
\bibinfo{author}{\bibfnamefont{P.~J.}
\bibnamefont{Kelly}}, and 
\bibinfo{author}{\bibfnamefont{F.~J.~A.}
    \bibnamefont{den~Broeder}}, \bibinfo{journal}{Phys. Rev. Lett.} \textbf{
\bibinfo{volume}{68}}, \bibinfo{pages}{682} (\bibinfo{year}{1992}).

\bibitem[Choy et~al.(2000)Choy, Naset, Chen, Hershfield, and Stanton]
{Choy:baps00} \bibinfo{author}{\bibfnamefont{T.-S.} \bibnamefont{Choy}}, 
\bibinfo{author}{\bibfnamefont{J.}~\bibnamefont{Naset}}, \bibinfo{author}{
\bibfnamefont{J.}~\bibnamefont{Chen}}, \bibinfo{author}{\bibfnamefont{S.}~\bibnamefont{Hershfield}}, and \bibinfo{author}{\bibfnamefont{C.}~\bibnamefont{Stanton}}, \bibinfo{journal}{Bull. Am. Phys.
Soc.} \textbf{\bibinfo{volume}{45}}, \bibinfo{pages}{L36}
(\bibinfo{year}{2000}), \bibinfo{note}{http://www.phys.ufl.edu/fermisurface/}.

\bibitem[Pajda et~al.(2001)Pajda, Kudrnovsk{\'{y}}, Turek, Drchal, and Bruno]
{Pajda:prb01} \bibinfo{author}{\bibfnamefont{M.}~\bibnamefont{Pajda}}, 
\bibinfo{author}{\bibfnamefont{J.}~\bibnamefont{Kudrnovsk{\'{y}}}}, 
\bibinfo{author}{\bibfnamefont{I.}~\bibnamefont{Turek}}, \bibinfo{author}{
\bibfnamefont{V.}~\bibnamefont{Drchal}}, and \bibinfo{author}{
\bibfnamefont{P.}~\bibnamefont{Bruno}}, \bibinfo{journal}{Phys. Rev. B} 
\textbf{\bibinfo{volume}{64}}, \bibinfo{pages}{174402} (\bibinfo{year}{2001}).

\bibitem[K{\"{u}}bler(2000)]{Kubler00} \bibinfo{author}{\bibfnamefont{J.}~\bibnamefont{K{\"{u}}bler}}, \emph{
\bibinfo{title}{Theory of Itinerant
Electron Magnetism}} (\bibinfo{publisher}{Oxford University Press}, 
\bibinfo{address}{Oxford}, \bibinfo{year}{2000}).

\bibitem[Mertig et~al.(1993)Mertig, Zeller, and Dederichs]{Mertig:prb93} 
\bibinfo{author}{\bibfnamefont{I.}~\bibnamefont{Mertig}}, 
\bibinfo{author}{\bibfnamefont{R.}~\bibnamefont{Zeller}}, and 
\bibinfo{author}{\bibfnamefont{P.~H.}
    \bibnamefont{Dederichs}}, \bibinfo{journal}{Phys. Rev. B} \textbf{
\bibinfo{volume}{47}}, \bibinfo{pages}{16178} (\bibinfo{year}{1993}).

\bibitem[Kunes and Kambersk{\'{y}}(2002)]{Kunes:prb02} \bibinfo{author}{
\bibfnamefont{J.}~\bibnamefont{Kunes}} and \bibinfo{author}{
\bibfnamefont{V.}~\bibnamefont{Kambersk{\'{y}}}}, 
\bibinfo{journal}{Phys.
Rev. B} \textbf{\bibinfo{volume}{65}}, \bibinfo{pages}{212411}
(\bibinfo{year}{2002}); \bibinfo{author}{\bibfnamefont{D.}~\bibnamefont{Steiauf}} and \bibinfo{author}{\bibfnamefont{M.}~\bibnamefont{F\"{a}hnle}}, \bibinfo{journal}{{\em ibid.}} \textbf{\bibinfo{volume}{72}}, 
\bibinfo{pages}{064450} (\bibinfo{year}{2005}); \bibinfo{author}{
\bibfnamefont{J.}~\bibnamefont{Ho}}, \bibinfo{author}{\bibfnamefont{F.~C.}
\bibnamefont{Khanna}}, and 
\bibinfo{author}{\bibfnamefont{B.~C.}
    \bibnamefont{Choi}}, \bibinfo{journal}{Phys. Rev. Lett.} \textbf{
\bibinfo{volume}{92}}, \bibinfo{pages}{097601} (\bibinfo{year}{2004}).

\bibitem[Koopmans et~al.(2005)Koopmans, Ruigrok, {Dalla Longa}, and de~Jonge]
{Koopmans:prl05} \bibinfo{author}{\bibfnamefont{B.}~\bibnamefont{Koopmans}}, 
\bibinfo{author}{\bibfnamefont{J.~J.~M.} \bibnamefont{Ruigrok}}, 
\bibinfo{author}{\bibfnamefont{F.}~\bibnamefont{{Dalla Longa}}}, and 
\bibinfo{author}{\bibfnamefont{W.~J.~M.}
    \bibnamefont{de~Jonge}}, \bibinfo{journal}{Phys. Rev. Lett.} \textbf{
\bibinfo{volume}{95}}, \bibinfo{pages}{267207} (\bibinfo{year}{2005}).

\bibitem[Yao et~al.(2004)Yao, Kleinman, MacDonald, Sinova, Jungwirth, Wang,
Wang, and Niu]{Yao:prl04} \bibinfo{author}{\bibfnamefont{Y.}~\bibnamefont{Yao}}, \bibinfo{author}{\bibfnamefont{L.}~\bibnamefont{Kleinman}}, 
\bibinfo{author}{\bibfnamefont{A.~H.}
\bibnamefont{MacDonald}}, \bibinfo{author}{\bibfnamefont{J.}~\bibnamefont{Sinova}}, \bibinfo{author}{\bibfnamefont{T.}~\bibnamefont{Jungwirth}}, 
\bibinfo{author}{\bibfnamefont{D.~S.}
\bibnamefont{Wang}}, \bibinfo{author}{\bibfnamefont{E.}~\bibnamefont{Wang}},
and \bibinfo{author}{\bibfnamefont{Q.}~\bibnamefont{Niu}}, 
\bibinfo{journal}{Phys. Rev. Lett.} \textbf{\bibinfo{volume}{92}}, 
\bibinfo{pages}{037204} (\bibinfo{year}{2004}).

\bibitem[Berger(1996)]{Berger:prb96} \bibinfo{author}{\bibfnamefont{L.}~\bibnamefont{Berger}}, \bibinfo{journal}{Phys. Rev. B} \textbf{\bibinfo{volume}{54}}, \bibinfo{pages}{9353} (\bibinfo{year}{1996}).

\bibitem[Bazaliy et~al.(1998)Bazaliy, Jones, and Zhang]{Bazaliy:prb98} 
\bibinfo{author}{\bibfnamefont{Y.~B.} \bibnamefont{Bazaliy}}, 
\bibinfo{author}{\bibfnamefont{B.~A.} \bibnamefont{Jones}}, and 
\bibinfo{author}{\bibfnamefont{S.-C.}
    \bibnamefont{Zhang}}, \bibinfo{journal}{Phys. Rev. B} \textbf{\bibinfo{volume}{57}}, \bibinfo{pages}{R3213} (\bibinfo{year}{1998}).

\bibitem[Fern\'{a}ndez-Rossier et~al.(2004)Fern\'{a}ndez-Rossier, Braun, N{\'{u}}{\~{n}}ez, and MacDonald]{Fernandez:prb04} \bibinfo{author}{\bibfnamefont{J.}~\bibnamefont{Fern\'{a}ndez-Rossier}}, \bibinfo{author}{
\bibfnamefont{M.}~\bibnamefont{Braun}}, 
\bibinfo{author}{\bibfnamefont{A.~S.}
    \bibnamefont{N{\'{u}}{\~{n}}ez}}, and \bibinfo{author}{
\bibfnamefont{A.~H.}  \bibnamefont{MacDonald}}, 
\bibinfo{journal}{Phys.
Rev. B} \textbf{\bibinfo{volume}{69}}, \bibinfo{pages}{174412} (\bibinfo{year}{2004}).

\bibitem[Tatara and Kohno(2004)]{Tatara:prl04} \bibinfo{author}{
\bibfnamefont{G.}~\bibnamefont{Tatara}} and \bibinfo{author}{
\bibfnamefont{H.}~\bibnamefont{Kohno}}, \bibinfo{journal}{Phys. Rev. Lett.} 
\textbf{\bibinfo{volume}{92}}, \bibinfo{pages}{086601} (\bibinfo{year}{2004}
).

\bibitem[Shibata et~al.(2005)Shibata, Tatara, and Kohno]{Shibata:prl05} 
\bibinfo{author}{\bibfnamefont{J.}~\bibnamefont{Shibata}}, 
\bibinfo{author}{\bibfnamefont{G.}~\bibnamefont{Tatara}}, and 
\bibinfo{author}{\bibfnamefont{H.}~\bibnamefont{Kohno}}, 
\bibinfo{journal}{Phys. Rev. Lett.} \textbf{\bibinfo{volume}{94}}, 
\bibinfo{pages}{076601} (\bibinfo{year}{2005}).

\bibitem[Li and Zhang(2004{b})]{Li:prl04} \bibinfo{author}{
\bibfnamefont{Z.}~\bibnamefont{Li}} and \bibinfo{author}{\bibfnamefont{S.}~\bibnamefont{Zhang}}, \bibinfo{journal}{Phys. Rev. Lett.} \textbf{
\bibinfo{volume}{92}}, \bibinfo{pages}{207203} (\bibinfo{year}{2004}{\natexlab{b}}); \bibinfo{journal}{Phys. Rev. B} \textbf{\bibinfo{volume}{70}}, \bibinfo{pages}{024417} (\bibinfo{year}{2004}{\natexlab{a}}).

\bibitem[Li et~al.(2005)Li, He, and Zhang]{Li:jap05} \bibinfo{author}{
\bibfnamefont{Z.}~\bibnamefont{Li}}, \bibinfo{author}{\bibfnamefont{J.}~\bibnamefont{He}}, and \bibinfo{author}{\bibfnamefont{S.}~\bibnamefont{Zhang}}, \bibinfo{journal}{J.  Appl. Phys.} \textbf{
\bibinfo{volume}{97}}, \bibinfo{pages}{10C703} (\bibinfo{year}{2005}).

\bibitem[Zhang and Li(2004)]{Zhang:prl04} \bibinfo{author}{
\bibfnamefont{S.}~\bibnamefont{Zhang}} and \bibinfo{author}{
\bibfnamefont{Z.}~\bibnamefont{Li}}, \bibinfo{journal}{Phys. Rev. Lett.} 
\textbf{\bibinfo{volume}{93}}, \bibinfo{pages}{127204} (\bibinfo{year}{2004}).

\bibitem[Barnes and Maekawa(2005)]{Barnes:prl05} \bibinfo{author}{
\bibfnamefont{S.~E.} \bibnamefont{Barnes}} and \bibinfo{author}{
\bibfnamefont{S.}~\bibnamefont{Maekawa}}, 
\bibinfo{journal}{Phys. Rev.
Lett.} \textbf{\bibinfo{volume}{95}}, \bibinfo{pages}{107204} (\bibinfo{year}{2005}).

\bibitem[Belzig et~al.(1999)Belzig]{Belzig:sm99} \bibinfo{author}{
\bibfnamefont{W.}~\bibnamefont{Belzig}}, \bibinfo{author}{
\bibfnamefont{F.~K.} \bibnamefont{Wilhelm}}, \bibinfo{author}{
\bibfnamefont{C.}~\bibnamefont{Bruder}}, \bibinfo{author}{\bibfnamefont{G.}~\bibnamefont{Sch{\"{o}}n}}, and 
\bibinfo{author}{\bibfnamefont{A.~D.}
    \bibnamefont{Zaikin}}, \bibinfo{journal}{Superl. Microstr.} \textbf{
\bibinfo{volume}{25}}, \bibinfo{pages}{1251} (\bibinfo{year}{1999}).

\bibitem[Kohno et~al.(2006)Kohno, Tatara and Shibata]{Kohno:condmat06} 
\bibinfo{author}{\bibfnamefont{H.}~\bibnamefont{Kohno}}, \bibinfo{author}{
\bibfnamefont{G.}~\bibnamefont{Tatara}}, and \bibinfo{author}{
\bibfnamefont{J.}~\bibnamefont{Shibata}}, \bibinfo{note}{
\emph{cond-mat/}0605186} (\bibinfo{year}{2006}).

\bibitem[Runge and Gross(1984)]{Runge:prl84} \bibinfo{author}{
\bibfnamefont{E.}~\bibnamefont{Runge}} and 
\bibinfo{author}{\bibfnamefont{E.~K.~U.}
    \bibnamefont{Gross}}, \bibinfo{journal}{Phys. Rev. Lett.} \textbf{
\bibinfo{volume}{52}}, \bibinfo{pages}{997} (\bibinfo{year}{1984}).

\bibitem[Capelle et~al.(2001)Capelle, Vignale, and Gy{\"{o}}rffy]
{Capelle:prl01} \bibinfo{author}{\bibfnamefont{K.}~\bibnamefont{Capelle}}, 
\bibinfo{author}{\bibfnamefont{G.}~\bibnamefont{Vignale}}, and 
\bibinfo{author}{\bibfnamefont{B.~L.}
    \bibnamefont{Gy{\"{o}}rffy}}, \bibinfo{journal}{Phys. Rev. Lett.} 
\textbf{\bibinfo{volume}{87}}, \bibinfo{pages}{206403} (\bibinfo{year}{2001}).

\bibitem[Qian and Vignale(2002)]{Qian:prl02} \bibinfo{author}{
\bibfnamefont{Z.}~\bibnamefont{Qian}} and \bibinfo{author}{
\bibfnamefont{G.}~\bibnamefont{Vignale}}, 
\bibinfo{journal}{Phys. Rev.
Lett.} \textbf{\bibinfo{volume}{88}}, \bibinfo{pages}{056404} (\bibinfo{year}{2002}).

\bibitem[Rammer and Smith(1986)]{Rammer:rmp86} \bibinfo{author}{
\bibfnamefont{J.}~\bibnamefont{Rammer}} and \bibinfo{author}{
\bibfnamefont{H.}~\bibnamefont{Smith}}, \bibinfo{journal}{Rev. Mod. Phys.} 
\textbf{\bibinfo{volume}{58}}, \bibinfo{pages}{323} (\bibinfo{year}{1986}).

\bibitem[Valet and Fert(1993)]{Valet:prb93} \bibinfo{author}{
\bibfnamefont{T.}~\bibnamefont{Valet}} and \bibinfo{author}{
\bibfnamefont{A.}~\bibnamefont{Fert}}, \bibinfo{journal}{Phys. Rev. B} 
\textbf{\bibinfo{volume}{48}}, \bibinfo{pages}{7099} (\bibinfo{year}{1993}).

\bibitem[Ingvarsson et~al.(2002)Ingvarsson, Ritchie, Liu, Xiao, Slonczewski,
Trouilloud, and Koch]{Ingvarsson:prb02} \bibinfo{author}{\bibfnamefont{S.}~\bibnamefont{Ingvarsson}}, \bibinfo{author}{\bibfnamefont{L.}~\bibnamefont{Ritchie}}, 
\bibinfo{author}{\bibfnamefont{X.~Y.}
\bibnamefont{Liu}}, \bibinfo{author}{\bibfnamefont{G.}~\bibnamefont{Xiao}}, 
\bibinfo{author}{\bibfnamefont{J.~C.} \bibnamefont{Slonczewski}}, 
\bibinfo{author}{\bibfnamefont{P.~L.} \bibnamefont{Trouilloud}}, and 
\bibinfo{author}{\bibfnamefont{R.~H.}
    \bibnamefont{Koch}}, \bibinfo{journal}{Phys. Rev. B} \textbf{
\bibinfo{volume}{66}}, \bibinfo{pages}{214416} (\bibinfo{year}{2002}).

\bibitem[Tserkovnyak et~al.(2004)Tserkovnyak, Fiete, and Halperin]
{Tserkovnyak:apl04} \bibinfo{author}{\bibfnamefont{Y.}~\bibnamefont{Tserkovnyak}}, 
\bibinfo{author}{\bibfnamefont{G.~A.}
\bibnamefont{Fiete}}, and 
\bibinfo{author}{\bibfnamefont{B.~I.}
    \bibnamefont{Halperin}}, \bibinfo{journal}{Appl. Phys. Lett.} \textbf{
\bibinfo{volume}{84}}, \bibinfo{pages}{5234} (\bibinfo{year}{2004}).

\bibitem[Tserkovnyak et~al.(2005)Tserkovnyak]{Tserkovnyak:rmp05} 
\bibinfo{author}{\bibfnamefont{Y.}~\bibnamefont{Tserkovnyak}}, 
\bibinfo{author}{\bibfnamefont{A.}~\bibnamefont{Brataas}}, 
\bibinfo{author}{\bibfnamefont{G.~E.~W.} \bibnamefont{Bauer}}, and 
\bibinfo{author}{\bibfnamefont{B.~I.}
    \bibnamefont{Halperin}}, \bibinfo{journal}{Rev. Mod. Phys.} \textbf{
\bibinfo{volume}{77}}, \bibinfo{pages}{1375} (\bibinfo{year}{2005}).

\bibitem[Yamaguchi et~al.(2004)Yamaguchi, Ono, Nasu, Miyake, Mibu, and Shinjo
]{Yamaguchi:prl04} \bibinfo{author}{\bibfnamefont{A.}~\bibnamefont{Yamaguchi}}, \bibinfo{author}{\bibfnamefont{T.}~\bibnamefont{Ono}}, \bibinfo{author}{\bibfnamefont{S.}~\bibnamefont{Nasu}}, 
\bibinfo{author}{\bibfnamefont{K.}~\bibnamefont{Miyake}}, 
\bibinfo{author}{\bibfnamefont{K.}~\bibnamefont{Mibu}}, and 
\bibinfo{author}{\bibfnamefont{T.}~\bibnamefont{Shinjo}}, 
\bibinfo{journal}{Phys. Rev. Lett.} \textbf{\bibinfo{volume}{92}}, 
\bibinfo{pages}{077205} (\bibinfo{year}{2004}).
\end{thebibliography}
\end{document}